# Solutions of Schrödinger Equation with Generalized Inverted Hyperbolic Potential

Akpan N.Ikot*[1], Oladunjoye A.Awoga[1], Louis E.Akpabio[1] and Benedict I.Ita[2]

1Theoretical Physics group, Department of Physics,University of Uyo,Nigeria.

2Theoretical Quantum chemistry group,Department of Chemistry, University of calabar ,Nigeria.

*email: ndemikot2005@yahoo.com

## Abstract

We present the bound state solutions of the Schrödinger equation with generalized inverted hyperbolic potential using the Nikiforov-Uvarov method. We obtain the energy spectrum and the wave function with this potential for arbitrary $l$- state. We show that the results of this potential reduced to the standard known potentials - Rosen-Morse, Poschl . Teller and Scarf potential as special cases. We also discussed the energy equation and the wave function for these special cases.

# 1. Introduction

The analytical and numerical solutions of the wave equations for both relativistic and non-relativistic cases have taken a great deal of interest in recent times. In many cases different attempts have been developed to solve the energy eigenvalues from the wave equations exactly or numerically for non- zero angular momentum quantum number $(l \neq 0)$ for a given potential [1-6]. It is well known that these solutions play an essential role in the relativistic and non-relativistic quantum mechanics for some physical potentials of interest [7-10].

In this paper, we aim to solve the radial Schrödinger equation for quantum mechanical system with inverted generalized hyperbolic potential and show the results for this potential using Nikiforov-Uvarov method (NU)[11]. The present paper is an attempt to carry out the analytical solutions of the Schrodinger equation with the generalized inverted hyperbolic potential using the Nikiforov-Uvarov method.

The hyperbolic potentials under investigations are commonly used to model inter-atomic and intermolecular forces [3, 12]. Among such potentials are Poschl-Teller, Rosen-Morse and Scarf potential, which have been studied extensively in the literatures [13-20]. However, some of these hyperbolic potentials are exactly solvable or quasi-exactly solvable and their bound state solutions have been reported [21-27]. We seek to present and study a generalized hyperbolic potential which other potentials can be deduced as special cases within the framework of Schrödinger equation with mass m and potential V.

To the best as our knowledge no attempts have been reported that study the Schrödinger equation for generalized inverted hyperbolic potential. The paper is organized as follows:

Section II is devoted to the review of the Nikiforov-Uvarov method. In section III we present the exact solution of the Schrodinger equation. Discussion and results are presented in section IV. Finally we give a brief conclusion in section V.

**II Review of Nikiforov-Uvarov Method**

The NU method [11] was proposed and applied to reduce the second order differential equation to the hypergeometric . type equation by an appropriate co-ordinate transformation S = S(r) as [1-4,11].

$$\varphi''(s) + \frac{\bar{z}(s)}{\sigma(s)}\varphi'(s) + \frac{\bar{\sigma}(s)}{\sigma^2(s)}\varphi(s) = 0 \tag{1}$$

where $\sigma(s)$ and $\bar{\sigma}(s)$ are polynomials at most in the second order, and is a first . order polynomial. In order to find a particular solution of Eq. (1) we use the separation of variables with the transformation

$$\Psi(s) = \varphi(s)\chi(s) \tag{2}$$

It reduces Eq. (1) to an equation as hypergeometric type

$$\sigma(s)\chi''(s) + z(s)\chi'(s) + \lambda\chi(s) = 0 \tag{3}$$

and $\varphi(s)$ is defined as a logarithmic derivative in the following form and its solution can be obtained from

$$\frac{\varphi'(s)}{\varphi(s)} = \frac{\pi(s)}{\sigma(s)} \tag{4}$$

The other part of the wave formation $\chi(s)$ is the hypergeometric type function whose polynomial solutions are given by Rodriques relations.

$$\chi_n(s) = \frac{B_n}{\rho(s)} \frac{d^n}{ds^n}\left[\sigma^n(s)\rho(s)\right] \tag{5}$$

where $B_n$ is a normalization constant, and the weight function $\rho(s)$ must satisfy the condition.

$$\frac{d}{ds}(\sigma\rho) = \tau(s)\rho(s) \tag{6}$$

with

$$\tau(s) = \bar{\tau}(s) + 2\pi(s) \tag{7}$$

The function $\pi(s)$ and the parameter $\lambda$ requires for the NU method are defined as follows:

$$\pi(s) = \frac{\sigma' - \bar{\tau}}{2} \pm \sqrt{\left(\frac{\sigma' - \bar{\tau}}{2}\right)^2 - \bar{\sigma} + k\sigma} \tag{8}$$

$$\lambda = k + \pi'(s) \tag{9}$$

On the other hand in order to find the value of k, the expression under the square root of Eq. (8) must be square of polynomial. Thus, a new eigenvalue for the second order equation becomes

$$\lambda = \lambda_n = -n\frac{d\tau}{ds} - \frac{n(n-1)\sigma''}{2} \tag{10}$$

where the derivative $\frac{()}{}$ is negative. By comparison of Eqs. (9) and (10), we obtained the energy eigenvalues.

## III Bound State Solutions of the Schrödinger Equation

The Schrödinger equation with mass m and potential V(r) takes the following form [1-4]

$$\Psi''(r) + \frac{2m}{\hbar^2}[E - V(r)]\Psi(r) = 0 \tag{11}$$

where the generalized hyperbolic potential V(r) under investigation is defined as

$$V_{a,b,c,d}(r) = -aV_0 Coth(\alpha r) + bV_1 Coth^2(\alpha r) - cV_2 Co\sec h(\alpha r) + d \tag{12}$$

Here $V_0$, $V_1$ and $V_2$ are the depth of the potential and a, b, c and d are real numbers. The generalized hyperbolic potential V(r) of Eq.(12) has the following special cases:

(i)  $\quad V_{-a,0,c,0}(r) = aV_0 Coth(\alpha r) - cV_2 Co\sec h^2(\alpha r)$  $\hfill (13)$

(ii) $\quad V_{0,0,c,0}(r) = cV_2 Co\sec h^2(\alpha r)$  $\hfill (14)$

(iii) $\quad V_{0,b,0,0}(r) = bCoth^2(\alpha r)$  $\hfill (15)$

The potentials (13) . (15) are the Rosen-Morse potential, Poschl-Teller potential and Scarf potential respectively.

We now perform the transformation [2-4]

$$\Psi(r) = \frac{R(r)}{r} \qquad (16)$$

on equation (1) and obtain

$$R''(r) + \frac{2m}{\hbar^2}\left[E + aV_0 \coth(\alpha r) - bV_1 \coth^2(\alpha r) + cV_2 \operatorname{cosech}^2(\alpha r) - d\right]R(r) = 0 \qquad (17)$$

where the prime indicates differentiation both respect to r.

Now using a new ansaltz for the wave function in the form [21-23]

$$R(r) = e^{-\frac{\beta r}{2}} F(r) \qquad (18)$$

and including the centrifugal term, reduces Eq. (17) into the following differential equation,

$$\frac{d^2 R(r)}{dr^2} - \beta \frac{dF}{dr} + \frac{2m}{\hbar^2}[E + aV_O \coth(\alpha r) - bV_1 \coth^2(\alpha r)$$

$$+ cV_2 \operatorname{cosech}^2(\alpha r) - d - \frac{(l+1)}{r^2} + \left(\frac{\beta}{2}\right)^2]F(r) = 0 \qquad (19)$$

Because of the centrifugal term in Eq. (19), this equation cannot be solved analytically when the angular momentum quantum number $\ell \neq 0$. Therefore, in order to find the approximate analytical solution of Eq. (19) with $\ell \neq 0$, we must make an approximation for the centrifugal term. Thus, when $\alpha r \ll 1$ we use the approximation scheme [28-30] for the centrifugal term,

$$\frac{1}{r^2} \approx \alpha^2 \operatorname{cosech}^2(\alpha r) \qquad (20)$$

Substituting Eq. (20) into Eq. (19), we get

$$\frac{d^2F(r)}{dr^2} - \frac{\beta dF(r)}{dr} + \frac{2m}{\hbar^2}[E + \left(\frac{\beta}{\alpha}\right)^2 + aV_0 \coth(\alpha r) - bV_1 \coth^2(\alpha r)$$

$$+ cV_2 \cosech^2(\alpha r) - \alpha^2 l(l+1)\cosech^2(\alpha r) - d]F(r) = 0 \qquad (21)$$

Now making the change of variable

$$s = \coth(\alpha r) \qquad (22)$$

we obtain

$$(1+s^2)^2 \frac{d^2F}{ds^2} + (1+s^2)(\beta + 2s)\frac{dF}{ds}$$

$$+ \frac{2m}{\alpha^2 \hbar^2}\left[E + \left(\frac{\beta}{2}\right)^2 + aV_0 s - bV_1 s^2 + cV_2(1+s^2) - \alpha^2 l(l+1)(1+s^2) - d\right]F(s) = 0 \qquad (23)$$

Simplifying Eq. (23), we have

$$\frac{d^2F}{ds^2} + \frac{\beta + 2s}{(1+s^2)}\frac{dF}{ds} + \frac{1}{(1+s^2)^2}\left[-\varepsilon^2 + \beta^2 s + \gamma^2 s^2\right]F(s) = 0 \qquad (24)$$

where the following dimensionless parameters have been employed:

$$\varepsilon^2 = -\frac{2m}{\alpha^2 \hbar^2}\left[E + \left(\frac{\beta}{2}\right) + cV_2 - \alpha^2 \ell(\ell+1+d)\right],$$

$$\beta^2 = \frac{2maV_0}{\hbar^2 \alpha^2}$$

$$\gamma^2 = \frac{2m}{\hbar^2 \alpha^2}\left(cV_2 - bV_1 - \alpha^2 \ell(\ell+1)\right) \qquad (25)$$

Comparing Eq. (1) and Eq. (24), we obtain the following polynomials,

$$\tilde{\tau}(s) = (\beta + 2s), \quad \sigma(s) = (1+s^2), \quad \bar{\sigma}(s) = -\varepsilon^2 + \beta^2 s + \gamma^2 s^2 \qquad . \qquad (26)$$

Substituting these polynomials into Eq. (8), we obtain the $\pi(s)$ function as

$$\pi(s) = -\frac{\beta}{2} \pm \frac{1}{2}\sqrt{(4k-4\gamma^2)s^2 - 4\beta^2 s + \beta^2 + 4\varepsilon^2 + 4k} \qquad (27)$$

The expression in the square root of Eq. (27) must be square of polynomial in respect of the NU method. Therefore, we determine the ( )-values as

$$\pi(s) = -\frac{\beta}{2} \pm \frac{1}{2} \begin{cases} \sqrt{(u+v)}s + \sqrt{(u-v)} \\ \text{for } k = \gamma^2 - \varepsilon^2 - \left(\frac{\beta}{2}\right)^2 + \sqrt{u^2 - v^2} \\ \\ \sqrt{(u+v)}s - \sqrt{(u-v)} \\ \text{for } k = \gamma^2 - \varepsilon^2 - \left(\frac{\beta}{2}\right)^2 - \sqrt{u^2 - v^2} \end{cases} \qquad (28)$$

where $u = \left(\varepsilon^2 \sqrt{1 + \frac{\beta^2}{2\varepsilon^2}} + \gamma^2\right)$ and $V = i\beta\sqrt{\gamma^2 + \frac{5}{2}\beta^2}$

For the polynomial of $\tau = \bar{\tau} + 2\pi$ which has a negative derivative, we get

$$k = \gamma^2 - \varepsilon^2 - \left(\frac{\beta}{2}\right)^2 + \sqrt{u^2 - v^2} \qquad (29)$$

$$\pi(s) = -\frac{\beta}{2} - \frac{1}{2}\left[\sqrt{(u+v)}s - \sqrt{u-v}\right] \qquad (30)$$

Now using $\lambda = k + \pi'(s)$, we obtain $\tau(s)$ and $\lambda$ valued as

$$\tau(s) = 2s - \sqrt{(u+v)}s + \sqrt{u-v} \qquad (31)$$

$$\lambda(s) = \gamma^2 - \varepsilon^2 - \left(\frac{\beta}{2}\right)^2 - \sqrt{u^2 - v^2} - \frac{1}{2}\sqrt{u+v} \qquad (32)$$

Another definition of $\lambda_n$ is as given in Eq. (10), thus using values of $\tau(s)$ and $\sigma(s)$, we get,

$$\lambda = \lambda_n = u\sqrt{u+v} - u(u+1) \qquad (33)$$

Comparing Eq. (32) and (33), we obtain the energy eigenvalue equation as

$$\left[\frac{(n+1)}{8\sqrt{2}\beta\gamma}\right] + i\left(\frac{\gamma}{8\sqrt{2}\beta} - \frac{1}{2v}\right)\varepsilon^4 - \left[1 + \frac{i\beta^2}{4}\left(1+\frac{1}{v}\right)\right]\varepsilon^2$$

$$-\left[\Sigma - \frac{(n+1)}{2}\sqrt{v} + iv + \frac{\beta\gamma}{2\sqrt{2}}\left((n+1)+i\gamma^2\right) - \frac{i\gamma^4}{2}\right] = 0 \qquad (34)$$

where $\Sigma = \left(\frac{\beta}{2}\right)^2 - \gamma^2 - n(n+1)$

Solving the energy eiginvalue equation explicitly, we obtain the energy eiginvalues as

$$\varepsilon^2 = \frac{\left[1 + \frac{i\beta^2}{4}\left(1+\frac{1}{v}\right)\right]}{2\left[\frac{(n+1)}{8\sqrt{2}\beta r} + i\left(\frac{\gamma}{8\sqrt{2}\beta} - \frac{1}{2v}\right)\right]}$$

$$\pm \frac{1}{2\left[\frac{(n+1)}{8\sqrt{2}\beta\gamma} + i\left(\frac{\gamma}{8\sqrt{2}\beta} - \frac{1}{2v}\right)\right]}\sqrt{\left(1+\frac{i\beta^2}{4}\left(1+\frac{1}{v}\right)\right)^2 + 4\left[\frac{(n+1)}{8\sqrt{2}\beta\gamma} + i\left(\frac{\gamma}{8\sqrt{2}\beta} - \frac{1}{2v}\right)\right]}$$

$$\times \left[\Sigma - \frac{(n+1)\sqrt{v}}{2} + iV + \frac{\beta\gamma}{2\sqrt{2}}\left((n+1)+i\gamma^2\right) - \frac{i\gamma^4}{2}\right] \qquad (35)$$

Now using these quantities of Eq. (20) and the definition for $\Sigma$ and V given as

$$\Sigma = \frac{2m}{\hbar^2\alpha^2}\left[\frac{av_0}{2} - cV_2 + bV_1 + \alpha^2 l(\ell+1) + \frac{\hbar^2\alpha^2 n(n+1)}{2m}\right] \quad (36)$$

$$V = i\left(\frac{2m}{\hbar^2\alpha^2}\right)\sqrt{aV_0 + cV_2 - bV_1 - \alpha^2\ell(\ell+1)} \quad (37)$$

We obtain the energy spectrum of the Eq. (35) for the Schrödinger equation with the generalized inverted hyperbolic potential as

$$E_{nl} = \frac{-\frac{\hbar^2\alpha^2}{2m}\left[1 + \frac{i\beta^2}{4}(1+v)\right]}{2\left[\frac{(n+1)}{8\sqrt{2}\beta r} + i\left(\frac{r}{8\sqrt{2}\beta} - \frac{1}{2v}\right)\right]}$$

$$\pm \frac{\frac{\hbar^2\alpha^2}{2m}}{2\left[\frac{(n+1)}{8\sqrt{2}\beta r} + i\left(\frac{r}{8\sqrt{2}\beta} - \frac{1}{2v}\right)\right]}\sqrt{\left(1 + \frac{i\beta^2}{4}\left(1+\frac{1}{v}\right)\right)^2 + 4\left[\frac{(n+1)}{8\sqrt{2}\beta r} + i\left(\frac{r}{8\sqrt{2}\beta} - \frac{1}{v}\right)\right]} \times \left[\Sigma - \frac{(n+1)}{2}\sqrt{v} + iv + \frac{\beta r}{2\sqrt{2}}(n+1+ir^2) - \frac{ir^4}{2}\right]$$

$$-\left(\frac{\beta}{2}\right)^2 - cV_2 + \alpha^2\ell(\ell+1) - d \quad (38)$$

We now find the corresponding eigenfunctions. The polynomial solutions of the hyperbolic function $\chi_n(s)$ depend on the determination of weight function $\rho(s)$. Thus we determine the $\rho(s)$ function in Eq. (6) as

$$\rho(s) = (1+s^2)^{-2}\left(\frac{1+is}{1-is}\right)^{\mu+iv} \quad (39)$$

where $\mu = 2 - \sqrt{u+v}$, $\upsilon = \sqrt{u-v}$ and substituting Eq. (39) in to the Rodriques relation of Eq. (5), we have

$$\chi_n(s) = B_n(1+is)^{2+\mu+2\upsilon}(1-is)^{2-\mu-i\upsilon}$$

$$\times \frac{d^n}{ds^n}\left[(1+is)^{N+\mu+i\upsilon-2}(1-is)^{n-\upsilon-i\upsilon-2}\right] \quad (40)$$

The polynomial solution of $\chi_n(s)$ can be expressed in terms of Jacobi polynomials which is one of the orthogonal polynomials, which is $\rho_n^{2+A,2-A}(x)$, where $A = \mu + i\upsilon$, and $x = is$

The other part of the wave function is obtain from Eq. (4) as,

$$\varphi(s) = (1+is)^{\frac{\mu+B}{2}}(1-is)^{\frac{\mu-B}{2}} \quad (41)$$

where $B = \frac{\upsilon + \beta}{2i}$.

Combining the Jacobi polynomials of Eq. (40) and Eq. (41), we obtain the redial wave function of the Schrödinger equation with inverted generalized hyperbolic potential as

$$F_{nl}(S) = N_n(1+x)^{\frac{\mu+B}{2}}(1-x)^{\frac{\mu-B}{2}} P_n^{2+A,2-A}(x) \quad (43)$$

where $N_n$ is a new normalization constant and obeys the condition $\int_{-\infty}^{\infty} R_n^2(s)ds = 1$. The total radial wave function is obtain using Eq. (18) and Eq. (22) as

$$R_{nl}(r) = N_n[1+i\coth(\alpha r)]^{\frac{\mu+B}{2}}(1+i\coth(\alpha r))^{\frac{\mu-B}{2}} P_n^{2+A,2-A}(i\coth(\alpha r)) \quad (43)$$

## IV. Results and Discussion

The well-known potentials are obtained from the generalized inverted hyperbolic potential if we make appropriate choose for the values of the parameters in the generalized inverted potentials as stated in section III. We plotted the variation of the generalized inverted hyperbolic potential as a function of r for a = 1 , b = 0.01 , $V_0$ = 1Mev , $V_1$ = 0.5Mev , $C_2$ =2 , $V_2$ = 0.02Mev , d =2Mev at different parameters of  = 1,2,3 and 4 as display in figure 1

### IV.1 Rosen – Morse Potential

For b = d = 0, the Rosen . Morse Potential is obtain as given in Eq. (13). We plotted the variation of Rosen . Morse V(r) with r for a = -1 , $V_0$ = 1Mev , c =2   and $V_2$ = 0.02Mev  with different α parameters of α =1, 2, 3, and 4 in figure 2.  Substituting b = d =0 in Eq. (38) and Eq. (43), we obtain the energy spectrum and the wave function of the Rosen-Morse potential respectively.

### IV.2 Poschl – Teller Potential

Poschl-Teller Potential is obtain from the generalized inverted hyperbolic potential by setting a = b = d = 0 and c = -c as given in Eq. (14). The Poschl . Teller potential is plotted as a function of r for c = -2 and $V_2$ = 0.02Mev in fig. 3.  Substituting these parameters in the energy equation of Eq. (38) and wave function (43), we obtain the desire energy spectrum and the wave function of the Poschl . Teller potential.

### IV.3 Scarf Potential

We can deduce the Scarf potential from the generalized inverted hyperbolic potential by setting a =c = d = 0. We display in fig. 4 the plot of Scarf potential as a function of r for b = 0.05, $V_1$ =0.5Mev with various parameter of $\alpha$ = 1, 2, 3, and 4. Setting the above limiting values in Eq. (38) and Eq. (43) we obtain the energy eigen-values and wave function for the Scarf potential respectively.

## V. Conclusions

The bound state solutions of the Schrödinger equation with a generalized inverted hyperbolic potential have been investigated within the framework of the Nikiforov-Uvarov method. Three well-known potential have been deduced from this potential. We discussed the energy spectrum and the wave function of the SE with this potential for an arbitrary $l$- state. We also discussed the special cases of the generalized inverted hyperbolic potential: Rosen Morse, Poschl-Teller and Scarf potentials. Finally, we plotted the effective potential as a function of r for different l=1, 2 and3 as shown in fig.5

Acknowledgment

This work is partially supported by the Nandy Reserch Grant No. 64 . 01-2271

# References

1. A.N.Ikot, L.E.Akpabio and E.J.Uwah, EJTP, 8, 25(2011)225-232

2. A.N.Ikot, A.D.Antia, L.E.Akpabio and J.A.Obu, JVR, 6, 2(2011)65-76.

3. A.N.Ikot, L.E.Akpabio and J.A.Obu, JVR, 6, 1(2011)1-13.

4 R.Sever, C.Tezcan,O.Yesiltas and M.bucurget,Int.J.Theor.Phys.47(2007)2243-2248.

5. G.F.Wei, X. Y Liu and W. L Chen, Int. J Theor. Phys, 48, (2009) 1649-1658.

6. K. J Oyewumi, E. O Akinpelu and A. D Agboola, Int. J. Theor. Phys. 47, (2008), 1039-1057.

7. O. M Al-Dossary, Int. Jour. Quant. Chem, 107, 10 (2007), 2040-2046.

8. S. Ikhdair and R. Sever, Int. J. theor. Phys. 46 (2002) 2384-2395.

9. H. X. Quan, L. Guang, W. Z. Min, N. L. Bin and M. Yin, Commun. Theor. Phys. 53 (2010).

10. Y. F. Cheng and T. Q Dai, Chinese. J. Phys. 45, 5 (2007) 480-487

11. A. F. Nikiforov and U. B. Uvarov, Special Functions of Mathematical Physics, Birkhausa, Basel (1988)

12. A. S. Halberg, Int. J. Maths. And Mathematical Sciences, 2011, (2011) 1-9, DOI:10.1155/2011/358198

13. S. Meyur, EJTP, 8, 25 (2011) 217-224

14. S. G Roy, J. Choudhury, N. K. Sarkar, S. R. Karumuri and R. Bhattacharjee, EJTP, 7, 24 (2010) 230-240

15. S. Meyur and S. Debnath, Bul. J. Phys. 36, 1, (2009)

16. . S. Meyur and S. Debnath, Bul. J. Phys. 30, 4, (2009)

17. J. J Diaz, J. Negro, I. M Nieto and O. Rosan Ortizi, J. Phys A 32, (2007) 8447-8460

18. A. Contreras and D. J Fernandez, J. Phys. A 41, (2008) 475303

19. D. J Fernandez, arxiv: 0910.0192v1 (2009)

20. F. Corper, A. Khare and U. Sukhature, Supersymmetry and Quantum Mechanics, Phys. Rep. 251 (1990) 267-385

21. S. D. Hernandez and D. J Fernandez, Int. J. Theor. Phys. DOI: 10.10007/s 10773.010.05222

22. C. B. Compean and M. Kirchbach, Euro. Phys. J. A, 33 (2007) 1-4

23. C. B. Compean and M. Kirchbach, J. Phys. A 39 (2007) 547-557

24. A. N Ikot and L. E Akpabio, Appl. Phys. Res. 2, 2 (2010) 202-208

25. K. J Oyewumi and C. O Akoshile, Euro. Phys. J. A, 45 (2010) 311-318

26. W. A Yahya, K. J Oyewumi, C. O Akoshile and T. T Ibrahim, JVR 5, 3 (2010) 27-34

27. K. J Oyewumi, Found.Phys.Lett.18,75(2005)

28. Y. Xu, S. He and C. S Jia, Phys. Scr. 81 (2010) 045001/

29. R. I Greene and C Aldrich, Phys. Rev. A 14, 2363 (1976)

30. C. S Jia, T. Chen and L. G Cui, Phys. Lett. A 373, 162, (2009)

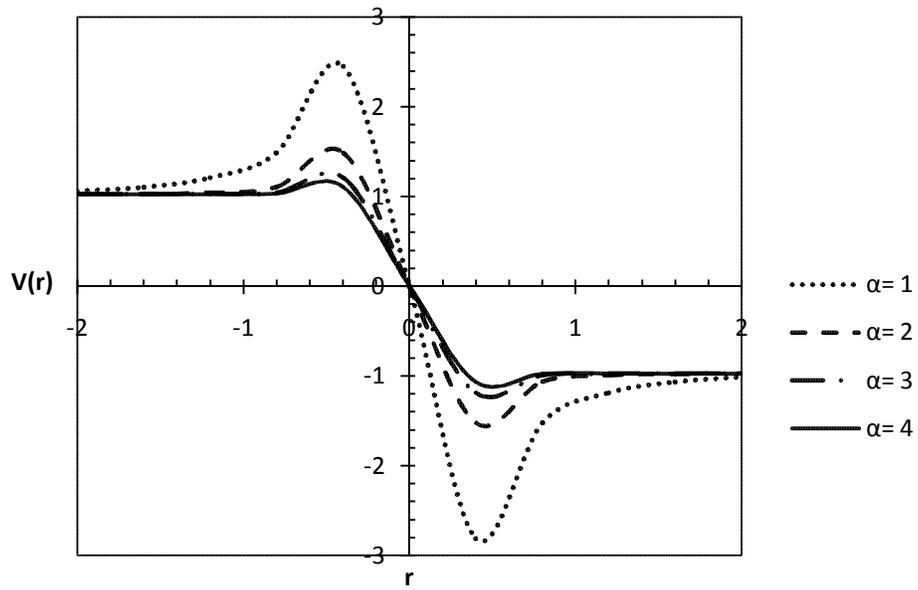

**Fig 1.** A plot of inverted generalized hyperbolic potential with r for a=1, 0.01, c=2, d=0.02, $V_0$=1Mev, $V_1$=0.5Mev, $V_2$=0.02MeV and α=1, 2, 3 and 4

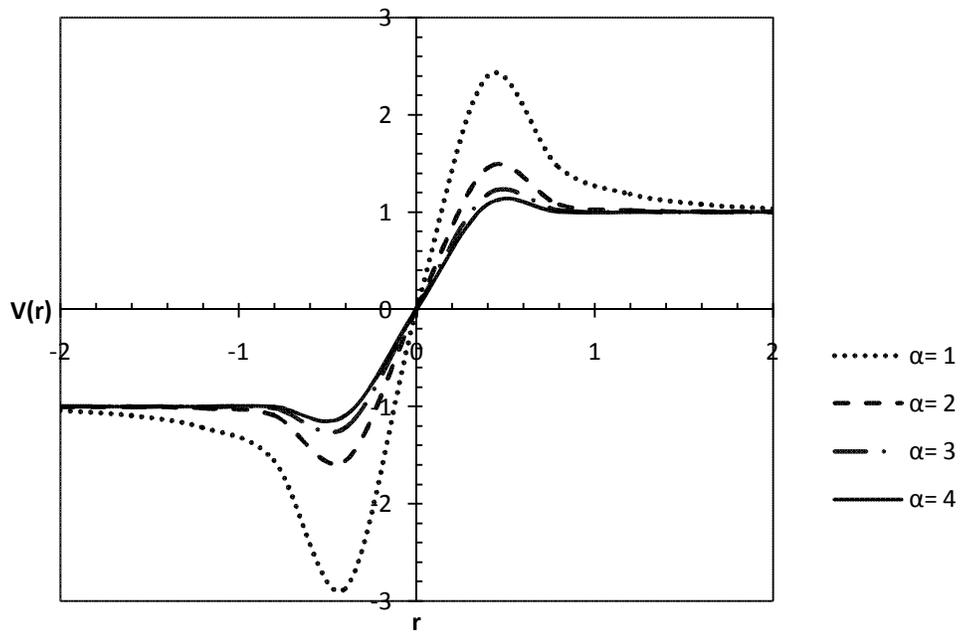

**Fig 2.** Variation of Rosen-Morse potential with r for a= -1, b=0, c=2, d=0, $V_0$=1MeV, $V_1$=0.5MeV, $V_2$=0.02MeV with various parameter of α=1, 2, 3 and 4

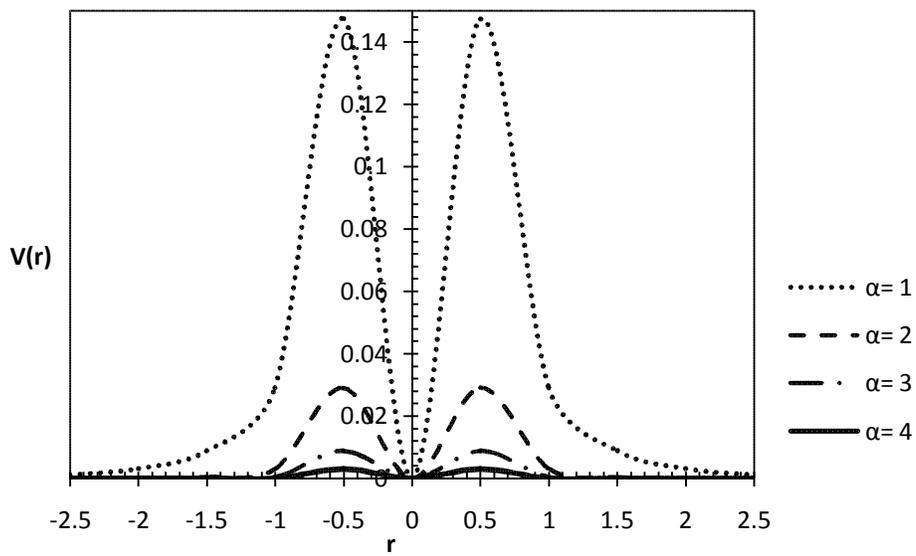

**Fig 3. A plot of Poschl-Teller potential with r for a= 0, b=0, c= -2, d=0, $V_0$=1MeV, $V_1$=0.5MeV , $V_2$=0.02MeV with various parameter of α=1, 2, 3 and 4**

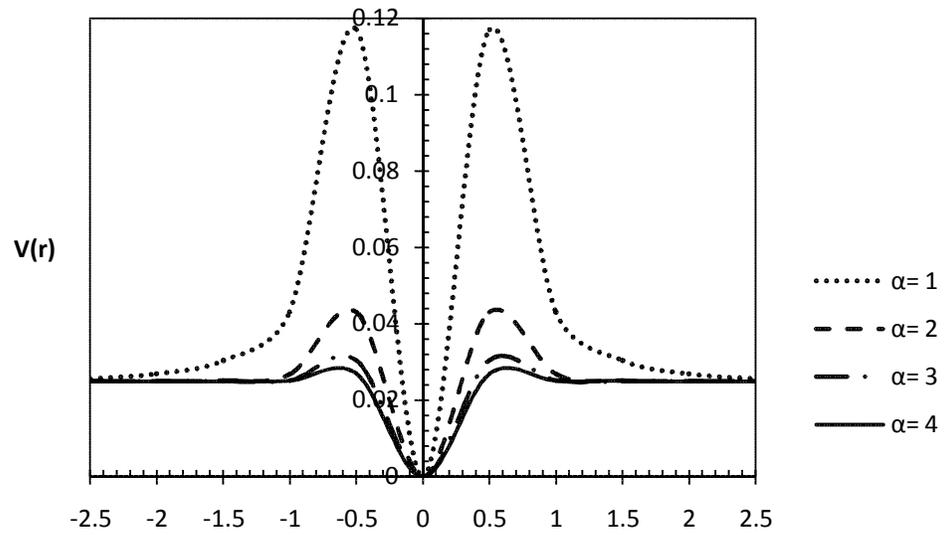

**Fig 4.** A plot of Scarf potential with r for a= 0, b=0.05, c=0, d=0, $V_0$=1MeV $V_1$=0.5MeV , $V_2$=0.02MeV with various parameter α=1, 2, 3 and 4

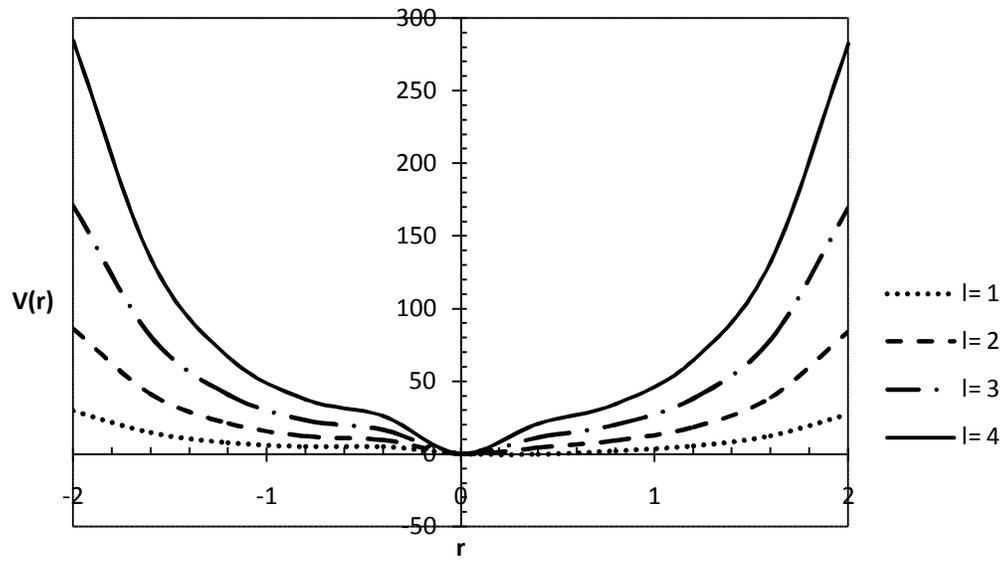

**Fig 5. A variation of the effective potential as a function of r for l=1, 2, 3 and 4 with α=1**